\documentclass[nofootinbib,notitlepage,superscriptaddress,10pt,aps,pra,twocolumn]{revtex4-1}

\usepackage[utf8x]{inputenc}
\usepackage{amsmath}
\usepackage{amssymb}
\usepackage{bbm}
\usepackage{graphicx}

\begin{document}

\begin{flushleft}
{\footnotesize Physical Review D 87, 084023 (2013)}\\
{\footnotesize http://link.aps.org/doi/10.1103/PhysRevD.87.084023}
\end{flushleft}

\title{Predictive description of 
Planck-scale-induced spacetime fuzziness}

\author{Giovanni AMELINO-CAMELIA}
\affiliation{Dipartimento di Fisica, Universit\`a di Roma ``La Sapienza", P.le A. Moro 2, 00185 Roma, EU}
\affiliation{INFN, Sez.~Roma1, P.le A. Moro 2, 00185 Roma, EU}

\author{Valerio ASTUTI}
\affiliation{Dipartimento di Fisica, Universit\`a di Roma ``La Sapienza", P.le A. Moro 2, 00185 Roma, EU}
\affiliation{INFN, Sez.~Roma1, P.le A. Moro 2, 00185 Roma, EU}

\author{Giacomo ROSATI}
\affiliation{Dipartimento di Fisica, Universit\`a di Roma ``La Sapienza", P.le A. Moro 2, 00185 Roma, EU}
\affiliation{INFN, Sez.~Roma1, P.le A. Moro 2, 00185 Roma, EU}
\affiliation{Institute for Theoretical Physics, University of Wroc\l{}aw, Pl.\ Maksa Borna
9, Pl--50-204}

\begin{abstract}
Several approaches to the quantum-gravity problem
predict that spacetime should be ``fuzzy", but have been so far unable to
provide a crisp physical characterization of this notion.
An intuitive picture of spacetime fuzziness has been proposed on the basis
of semi-heuristic arguments, and in particular involves an irreducible Planck-scale contribution
to the uncertainty of the energy of a particle.
These arguments also inspired a rather active phenomenological programme
looking for blurring of images of distant astrophysical sources that would result
from such energy uncertainties.
We here report the first ever physical
characterization of spacetime fuzziness derived constructively within a quantum picture of spacetime,
the one provided by spacetime noncommutativity.
Our results confirm
earlier heuristic arguments suggesting that
spacetime fuzziness, while irrelevantly small on terrestrial scales,
could be observably large for propagation
of particles over cosmological distances.
However, 
we find no Planck-scale-induced lower bound on the uncertainty of the energy of
particles, and we observe that this changes how we should picture a quantum spacetime
and also imposes a reanalysis of the associated phenomenology.
\end{abstract}

\maketitle

There has been growing interest~\cite{ngPRL,steingbring,shviliBLURRING,tamburini,ngARXIV2011}
in the possibility of testing the hypothesis of
spacetime fuzziness at the Planck scale ($E_P \sim 10^{28}eV$)
exploiting an associated effect blurring the images of
distant astrophysical sources, such as quasars.
The arguments providing encouragement for these phenomenological studies are merely heuristic,
but this could be a rare opportunity~\cite{carlipREVIEW,gacQM100,gacPRL2009} for testing experimentally
an aspect of the interplay between gravitational and quantum-mechanical phenomena.
The scenario considered in Refs.~\cite{ngPRL,steingbring,shviliBLURRING,tamburini,ngARXIV2011}
(building on earlier analogous pictures, such as those in Refs.~\cite{gacgwi,ngfoam,lieuHillman,ragazzoni})
is centered on the possibility that
 the quantum-gravity contribution
  to the fuzziness of a particle's worldline
might grow with propagation distance, in such a way that, in spite of its ultra-microscopic characteristic
scale (which we assume~\cite{carlipREVIEW,gacQM100,gacPRL2009}
to be of the order of the Planck length $\ell \simeq 10^{-35}m$),
it could turn into a macroscopic effect for propagation over suitably large (cosmological) distances.

The most crucial aspect of the phenomenological proposals
is the
assumption~\cite{ngPRL,steingbring,shviliBLURRING,tamburini,ngARXIV2011,lieuHillman,ragazzoni}
that in a fuzzy/quantum spacetime
there should be an irreducible Planck-scale contribution
to the uncertainty of energies, governed by a law of the form
\begin{equation}
\delta E_{[\ell]} \sim \ell^\alpha ~ E^{1+\alpha}~,
\label{fuzzyMOMENTUM}
\end{equation}
with $\alpha$, assumed~\cite{ngPRL,steingbring,shviliBLURRING,tamburini,ngARXIV2011,lieuHillman,ragazzoni}
to take values between 1/2 and 1,
being the single parameter which should discriminate
in this respect among different proposals for a quantum spacetime.
[We shall consistently place an index ``$[\ell]$", as done in (\ref{fuzzyMOMENTUM}),
when reporting an estimate of the Planck-scale contribution to a given quantity.]

As first observed in Ref.~\cite{lieuHillman},
Eq.~(\ref{fuzzyMOMENTUM})
would in turn produce uncertainties in the phase velocity $\mathsf{v}_p=E/p$ and
in the group velocity $\mathsf{v}_g=dE/dp$, both of order $\delta E_{[\ell]}/ E$
and uncorrelated. This would imply that
as a wave propagates it will experience~\cite{lieuHillman} essentially random mismatches,
of order $\delta E_{[\ell]}/ E$,
between its phase velocity and its group velocity.
In turn one can then notice~\cite{lieuHillman} that during the propagation time $t_{prop}=D/\mathsf{v}_g$
the phase  should normally advance 
from its initial value by an amount $2\pi \mathsf{v}_p t_{prop}/\lambda$,
{\it i.e.} $2\pi (\mathsf{v}_p/\mathsf{v}_g) D/\lambda$ (denoting with $\lambda$ the wavelength).
The net result would be~\cite{ngPRL,steingbring,shviliBLURRING,tamburini,ngARXIV2011,lieuHillman,ragazzoni}
an uncertainty for the phase of a wave governed by a law of the type
\begin{equation}
\delta \phi_{[\ell]} = 2 \pi \frac{D^\beta}{\lambda^\beta} \,
 \delta \! \left[\frac{\mathsf{v}_p}{\mathsf{v}_g} \right]_{[\ell]}
 \sim  \frac{D^\beta}{\lambda^\beta} \frac{2 \pi}{E}\, \delta E_{[\ell]}~,
\label{phaselieu}
\end{equation}
where $\beta$ is an additional phenomenological parameter concerning whether
or not the random mismatches between phase  and  group velocity
should add coherently: if they add coherently~\cite{lieuHillman} then $\beta=1$, while according
to the most popular picture where they do not add coherently~\cite{ngPRL} one should
have $\beta = 1-\alpha$.

The debate revolving around the blurring of images of distant astros that would result
from this phase uncertainty has revolved exclusively around the $\beta$ parameter,
and therefore the hypotheses needed for coherent (versus incoherent) addition of 
phase/group velocity mismatches. Instead the aspect which is of primary interest for us here,
concerning the Planck-scale-induced energy
uncertainty $\delta E_{[\ell]}$
in (\ref{phaselieu}), and its description according to (\ref{fuzzyMOMENTUM}),
remained so far unchallenged, and provides the core feature of this whole research line.
This assumption is motivated in the relevant heuristic arguments by
essentially noticing that
operatively energy is a notion derived from spacetime observations, and with spacetime
being ``fuzzy" at the Planck scale one should, according
to Ref.~\cite{ngPRL,steingbring,shviliBLURRING,tamburini,lieuHillman},
inevitably get ``fuzzy energies".

Besides being, as we stressed, the key ingredient
of an active phenomenological program,
this link from spacetime fuzziness to irreducible contributions to energy uncertainty could be a 
very significant
characterization of the quantum-gravity realm.
But its only basis are indeed heuristic arguments.
Even in the most studied formalizations of quantum properties of spacetime 
at the Planck scale, such as
 Loop Quantum Gravity~\cite{rovelliLivingReview} and
 spacetime noncommutativity~\cite{doplicherFR,majrue}, spacetime fuzziness has been so far
 characterized only at a rather formal level, unsuitable for phenomenology
 and inconclusive for what concerns the description of energy uncertainties.

We here report significant progress in formalizing and analyzing worldline fuzziness
within the quantum-spacetime framework of spacetime noncommutativity.
And in particular we establish some severe limitations to the applicability of the
arguments summarized above, suggesting that in a quantum spacetime
there should be an irreducible
 Planck-scale contribution
to energy uncertainties.
We do this by considering the two most studied examples of spacetime noncommutativity,
the case of Moyal noncommutativity~\cite{szabo} and the case 
of  $\kappa$-Minkowski noncommutativity~\cite{majrue,lukieANNALS}.
For reasons that shall be clear in light of the outcome of our analyses,
we first specialize to the case of a 2D (1+1-dimensional) $\kappa$-Minkowski spacetime,
so that the noncommutativity of coordinates is fully specified by
\begin{equation}
[x_1 , x_0]=i \ell x_1 ~.
\label{kappamink2D}
\end{equation}
We are evidently working
in units such that the speed-of-light scale $c$ and the Planck scale $\hbar$ are set to unity,
 and most of our results are
derived at leading order in $\ell$, which suffices for the purposes of the relevant phenomenology.

The starting point of this analysis 
is provided by our previous study
 in Ref.~\cite{fuzzy1pap}, where
we addressed one of the challenges which had obstructed the
characterization of worldline fuzziness in 
noncommutative spacetimes.
In $\kappa$-Minkowski the time coordinate is a noncommutative observable,
whereas in the standard formulation of quantum mechanics the time coordinate is merely an evolution
parameter.
This difficulty can be circumvented~\cite{fuzzy1pap}
 by resorting to results~\cite{halliwellQM,gambiniportoQM,rovellireisenb}
establishing a covariant formulation of quantum mechanics,
where both the spatial coordinates and the time coordinate
play the same type of role. 
Spatial and time coordinates are well-defined operators on
a ``kinematical Hilbert space", which is just an ordinary Hilbert space of normalizable
wave functions~\cite{rovellireisenb}. Observable features of the quantum theory
are coded  on
the ``physical Hilbert space", obtained from the kinematical Hilbert space
 by enforcing the on-shellness constraint (this constraint codifies dynamics
 in the same sense familiar for the covariant formulation of classical mechanics;
see, {\it e.g.}, chapter 4 of Ref.~\cite{henneauxBOOK}).

Within this formulation of quantum mechanics
time and spatial coordinates of course commute among themselves but do not commute with their conjugate momenta,
so that in particular in the 2D case one has~\cite{rovellireisenb}
\begin{equation}
\!\! [{\pi}_0,{q}_0] = i \,,~~[{\pi}_1 , {q}_1] = -i  \,,~~[{\pi}_1,{q}_0] = [{\pi}_0 , {q}_1] = 0
\label{pregeomPHSPACE}
\end{equation}

We observed in Ref.~\cite{fuzzy1pap} that
the $\kappa$-Minkowski defining commutator (\ref{kappamink2D})
is satisfied
by posing a relationship between the $\kappa$-Minkowski coordinates and the phase-space
operators of the covariant formulation of quantum mechanics (the ones of Eq.~(\ref{pregeomPHSPACE}), here viewed
merely as formal auxiliary operators~\cite{fuzzy1pap})
of the following form:
\begin{eqnarray}
x_1 = \, e^{\ell \pi_0} \, q_1 ~,~~~x_0 = q_0~.~~~\label{targetX}
\end{eqnarray}
And we also show in Ref.~\cite{fuzzy1pap} that the translational symmetries of $\kappa$-Minkowski
spacetime, which in terms of the $x_1,x_0$ coordinates require a rather sophisticated
formalization (see, {\it e.g.} Refs.~\cite{majrue,lukieANNALS,meljanacTRANSLATIONS}),
can be implemented in terms of the auxiliary variables  $q_0,q_1,\pi_0,\pi_1$
as standard translation transformations:
\begin{equation}
\!\!{\cal T}_{a^\mu} \triangleright f(x_0 , x_1) \equiv
f(q_0 ,  q_1 e^{\ell \pi_0 }) -i a^\mu \left[\pi_\mu ,f(q_0 ,  q_1 e^{\ell \pi_0 })\right]
\label{representationD}
\end{equation}

We also established~\cite{fuzzy1pap} that
the ``on-shellness operator" (the operator which, for massless particles,
should vanish on physical states)
can be written in terms of the auxiliary variables $\pi_1,\pi_0$
 as follows
\begin{equation}
 {\cal H}= \left(\frac{2}{\ell}\right)^{2}\sinh^{2}\left(\frac{\ell {\pi}_{0}}{2}\right) - e^{-\ell {\pi}_{0}} {\pi}_{1}^{2} ~.
\label{dalembertell}
\end{equation}

One more result of Ref.~\cite{fuzzy1pap} which is relevant for the observations we are here reporting concerns the measure for integration
over momenta, needed for evaluating scalar products when working in the ``momentum representation":
we found that the implementation of relativistic symmetries in
terms of the $\kappa$-Minkowski $x_1 , x_0$ coordinates induces
an $\ell$-deformed $\pi_0,\pi_1$-integration measure:  $d\pi_0 d\pi_1 \longrightarrow \exp (-\ell \pi_0)
d\pi_0 d\pi_1$.

In Ref.~\cite{fuzzy1pap} we only reached the point of analyzing
the kinematical Hilbert space. The form of the operator ${\cal H}$ was established,
 but we did not explore the implications of enforcing the Hamiltonian constraint
 in obtaining the physical Hilbert space.
 Since we are here interested in the fuzziness of the worldline of a physical particle
we must progress to that next level.
More precisely we characterize physical observables  of free relativistic quantum
particles in $\kappa$-Minkowski spacetime following the covariant prescription
adopted in Ref.~\cite{rovellireisenb}: we obtain the needed feature of
invariance of physical observables under the action of $\cal H$
by introducing a new scalar product~\cite{rovellireisenb}, that projects all the orbit
of the gauge transformation generated by $\cal H$ on the same state.
This allows to formally refer to states in the kinematical Hilbert space (but only as representatives
of an orbit) and for free massless particles leads to the study of scalar products of the 
type $\langle \psi|\phi\rangle_{\cal H} = \langle \psi|\delta\left({\cal H}\right)\Theta(\pi_0)|\phi\rangle$,
where $\Theta(\pi_0)$  specifies a restriction~\cite{rovellireisenb}
 to positive-energy solutions of the on-shellness constraint.
Accordingly in the ``momentum-space representation" one has 
\begin{eqnarray}
\langle \psi|\phi\rangle_{\cal H} &=& \int e^{-\ell \pi_0}d\pi_1 d\pi_0  \delta\left({\cal H}\right) \Theta(\pi_0) \psi^*(\pi)\phi(\pi)
 \nonumber
\end{eqnarray}
%

We focus here on the case of a localized massless particle, describable in terms
of a gaussian state\footnote{Of course in the massless-particle limit, here of interest,
one must proceed cautiously: $\Psi_{\overline{q}_0,\overline{q}_1} (\pi_\mu; \overline{\pi}_\mu , \sigma_\mu)$
must replaced by $\Psi^{\alpha}_{\overline{q}_0,\overline{q}_1} (\pi_\mu; \overline{\pi}_\mu , \sigma_\mu) =
 \exp (-\alpha/\pi_0^2) \Psi_{\overline{q}_0,\overline{q}_1} (\pi_\mu; \overline{\pi}_\mu , \sigma_\mu)$
 with alpha a small infrared regulator which never actually matters in the results
 we here exhibit.}
\begin{equation}
\Psi_{\overline{q}_0,\overline{q}_1} (\pi_\mu; \overline{\pi}_\mu , \sigma_\mu)
\!=\!\! N e^{- \frac{\left(\pi_0 - \overline{\pi}_0\right)^2}{4\sigma_0^2} - \frac{\left(\pi_1 - \overline{\pi}_1\right)^2}{4\sigma_1^2}} \!\! e^{ i \pi_0 \bar{q}_0 - i \pi_1 \bar{q}_1},
\nonumber
\end{equation}
where $N$ is a normalization constant
$$N^{-2}= \int e^{-\ell \pi_0}d\pi_1 d\pi_0  \delta\left({\cal H}\right) \Theta(\pi_0) |\Psi_{\overline{q}_0,\overline{q}_1} (\pi_\mu; \overline{\pi}_\mu , \sigma_\mu)|^2$$
and $\Psi_{\overline{q}_0,\overline{q}_1}$ is evidently written in
the momentum-space representation, with parameters
 $\overline{\pi}_0,\overline{\pi}_1,\sigma_0,\sigma_1,\overline{q}_0,\overline{q}_1$
 (with $\overline{q}_0,\overline{q}_1$
 highlighted, in the notation $\Psi_{\overline{q}_0,\overline{q}_1}$,
 since the issue of localization of the particle is predominantly
 connected with those two parameters).

Our $\Psi_{\overline{q}_0,\overline{q}_1}$ gives a state on our physical Hilbert space
of relativistic free-particle quantum mechanics, so it
identifies a fuzzy worldline~\cite{rovellireisenb}, as it shall be evident
also from what follows.
The expectation in $\Psi_{\overline{q}_0,\overline{q}_1}$
of the measurable quantity
described  by a self-adjoint operator ${\cal O}$ is computed
in terms of $\langle \Psi_{\overline{q}_0,\overline{q}_1}|{\cal O}|\Psi_{\overline{q}_0,\overline{q}_1}\rangle_{\cal H}$.

The next hurdle we must face concerns the identification of a well-defined observable
suitable for the characterization of the fuzziness of the worldline. The apparently
obvious choices, $x_1$ and $x_0$, are actually not suitable for this task, since they
are not self-adjoint operators on our physical Hilbert space (in particular they do not commute
with ${\cal H}$).
We propose to remedy this by focusing
on the following ``intercept operator" ${\cal A}$:
\begin{equation}
{\cal A}=e^{\ell \pi_0} \left( q_1 - {\cal V} q_0 - \frac{1}{2}[q_0,{\cal V}] \right)
\label{calAeq}
\end{equation}
where ${\cal V}$ is short-hand for
${\cal V} \equiv ( {\partial {\cal H}}/{\partial \pi^0})^{-1}
 {\partial{\cal H}}/{\partial \pi^1}$.

One may notice that ${\cal A}$ is describable as an $\ell$-deformed Newton-Wigner operator~\cite{newtonwigner}.
And it is well known that within special-relativistic quantum mechanics there is no better
estimator of localization than the Newton-Wigner operator (it can only be questioned for localization comparable to
the Compton wavelength of the particle~\cite{newtonwigner}, but this merely conceptual limit of ideal localization
 is evidently irrelevant for the level of localization achieved by particle production at, {\it e.g.}, quasars).
For our purposes it is important to notice that
 ${\cal A}$ is a good observable on our physical Hilbert space
(self-adjoint, commuting with ${\cal H}$)
and evidently in the classical limit ${\cal A}$ reduces to
the intercept of the particle worldline with the $x_1$ axis.

\noindent
Let us focus, for conceptual clarity, on the analysis of
 the properties of ${\cal A}$ for the case of $\Psi_{0,0}$, {\it i.e.}
for $\overline{q}_0=0,\overline{q}_1=0$.
One then easily finds that
$$\langle \Psi_{0,0} |{\cal A}|\Psi_{0,0}\rangle_{\cal H} = 0$$
so this is a case where the particle intercepts the observer Alice in her origin.
The fact that this intercept is fuzzy reflects the fuzziness of the worldline
described by $\Psi_{0,0}$, and in particular the leading $\ell$-dependent contribution
to this fuzziness is characterized by
\begin{equation}
\delta {\cal A}^2_{[\ell]} = \left(\langle \Psi_{0,0} |{\cal A}^2|\Psi_{0,0}\rangle_{\cal H}\right)_{[\ell]}
 \approx \ell \langle \pi_0\rangle \sigma^{-2}/2
\label{fuzzyalice}
\end{equation}
where for simplicity we assumed (as we shall do throughout)  that $\sigma_1$ is small enough, in comparison
to $\sigma_0$, $\bar{\pi}_1$, to allow a saddle point approximation in the $\pi_1$ integration;
 then $\sigma$ (without indices) is
the effective gaussian width after the saddle point approximation
in $\pi_1$: $\sigma^{-2} \equiv \sigma_{1}^{-2}+<{\cal V}>^2 \sigma_{0}^{-2}$.

In our proposed interpretation of the formalism Eq.~(\ref{fuzzyalice})
gives the fuzziness of the worldline ``at Alice" (at the point of crossing the origin
of Alice's reference frame).
It is interesting to also consider the perspective of observers reached by the particle
at cosmological distances from Alice.
These observers are those who are connected to Alice by a pure translation,
so that for them the state of the particle is $\Psi_{a_0,a_1}$,
and are such that $< \!\! {\cal A} \!\! >=0$,
 {\it i.e.} $\langle \Psi_{a_0,a_1} |{\cal A}|\Psi_{a_0,a_1}\rangle_{\cal H} = 0$.
Finding these observers amounts to finding the translation parameters $a^0,a^1$
such
that $\langle \Psi_{0,0} |{\cal T}^{-1}{\cal A} {\cal T}|\Psi_{0,0}\rangle_{\cal H} = 0$,
where ${\cal T}$ is the one of Eq.~(\ref{representationD}). Of course, this leads
to a one-parameter family of solutions (the family of observers ``on the worldline"),
which unsurprisingly takes the form $a^1 =  \langle {\cal V}\rangle a^0$.

Crucial for us is that these obervers with
vanishing expectation value for the intercept have values of
the uncertainty in the intercept $\delta {\cal A}$ given by
\begin{eqnarray}
\delta {\cal A}^2_{[\ell]} &=&
\left(\langle \Psi_{a^0,\langle {\cal V}\rangle a^0}|{\cal A}^2 |\Psi_{a^0,\langle {\cal V}\rangle a^0}\rangle_{\cal H}\right)_{[\ell]} \approx \nonumber\\
&\approx & \left(\frac{\ell \langle \pi_0 \rangle}{2 \sigma^2}+ \ell^2 a_0^2 \sigma^2 \right)
\label{mainresult}
\end{eqnarray}
So we do have here a quantum-spacetime picture that fits within the intuition inspiring
spacetime-fuzziness phenomenology: one can in fact interpret
our observer Alice, the observer on  the worldline
for whom the fuzziness of the intercept takes the minimum value,
as the observer at the source (where the particle is produced),
and then the intercept of the particle worldline with
the origin of the reference frames of observers distant from Alice
(where the particle could be detected)
has bigger uncertainty.

However, our formalization 
provides a quantification of the relevant effects that differs from what
had been suggested heuristically.
A crucial aspect of these differences is uncovered
by evaluating the energy uncertainty $\delta E$. In Refs.~\cite{fuzzy1pap} we established that $\pi_0$
does have a standard role of energy for particles in our $\kappa$-Minkowski spacetime, and therefore
$\delta E$ is given by $\langle \Psi_{a^0,\langle {\cal V}\rangle a^0}|\pi_0^2
 |\Psi_{a^0,\langle {\cal V}\rangle a^0}\rangle_{\cal H} -
\langle \Psi_{a^0,\langle {\cal V}\rangle a^0}|\pi_0 |\Psi_{a^0,\langle {\cal V}\rangle a^0}\rangle^2_{\cal H}$
for which we find
\begin{equation}
\delta E^2 \simeq \sigma^2 - 2 \ell E \sigma^2
\label{mainresultE}
\end{equation}
Remarkably this shows that, contrary to what had been assumed 
on the basis of heuristic arguments, in our
quantum spacetime there is no irreducible Planck-scale contribution to 
energy uncertainties (since $\sigma$ can take unboundedly small values in our gaussian states).
This is perhaps our most significant result, which we feel has very strong implications for both the phenomenology
and the theoretical understanding of spacetime fuzziness. Phenomenologically, the fact
that we found
no irreducible Planck-scale contribution to 
energy uncertainties renders completely inapplicable to $\kappa$-Minkowski the experimental bounds
on spacetime fuzziness derived
in Refs.~\cite{ngPRL,steingbring,shviliBLURRING,tamburini,ngARXIV2011,lieuHillman,ragazzoni}.

And it is relatively easy to apply the strategy of analysis adopted above for $\kappa$-Minkowski
also to the other much studied noncommutative
spacetime, the one with Moyal noncommutativity.
Denoting with $X_\mu$ the coordinates of the Moyal spacetime one has~\cite{szabo} 
that 
$$[X_\mu , X_\nu]=i \ell^2 \theta_{\mu \nu}~,$$
where the dimensionless noncommutativity parameters $\theta_{\mu \nu}$ are coordinate independent.
Applying our approach centered on the manifestly-covariant formulation
of quantum mechanics to the Moyal case is indeed easier
than for the $\kappa$-Minkowski case, because of the simplicity
of the relationship between the Moyal 
coordinates and the phase-space
operators of the covariant formulation of quantum mechanics (the ones of Eq.~(\ref{pregeomPHSPACE}), here viewed again as formal auxiliary operators): by posing
\begin{eqnarray}
X_\mu = q_\mu + \ell^2 \frac{\theta_{\mu \nu}}{2} \, \pi^\nu ~.~~~\label{targetXmoyal}
\end{eqnarray}
the Moyal commutation relations are automatically satisfied. Comparison of this (\ref{targetXmoyal})
for the Moyal case
to the complexity of its non-linear $\kappa$-Minkowski counterpart (\ref{targetX}) already suggests 
how much simpler it is to adopt our strategy of analysis when considering Moyal noncommutativity. 
In particular,
the seed for our $\kappa$-Minkowski findings is in the associated description of translations, 
which according to (\ref{representationD}) are such that
\begin{equation}
x^\prime_0 = 
x_0 + a_0= 
q_0 + a_0~,~~
x^\prime_j = (q_j+a_j) e^{\ell \pi_0 } =
x_j + a_j e^{\ell \pi_0 }
\nonumber
\end{equation}
One can easily trace back to the nontrivial $e^{\ell \pi_0 }$ operatorial factor
for translations of spatial coordinates the source of the mechanism that gives
increasing fuzziness as the particle propagates. For the Moyal case
from (\ref{targetXmoyal}) it follows that under translations one simply has
\begin{eqnarray}
X^\prime_\mu = q_\mu +a_\mu
+ \ell^2 \frac{\theta_{\mu \nu}}{2} \, \pi^\nu = X_\mu + a_\mu ~,\nonumber
\end{eqnarray}
and in light of this it is easy to follow the steps of our $\kappa$-Minkowski analysis
adapting them to the Moyal case, ultimately finding
that there is no Planck-scale-induced increase of the fuzziness as the particle propagates.

This also explains why we chose to keep our primary focus on $\kappa$-Minkowski:
the Moyal case is simpler but does not even provide the starting ingredient of
the heuristic pictures of spacetime fuzziness we intended to investigate.
In $\kappa$-Minkowski we at least did find that, as argued by the heuristic arguments,
the fuzziness of a particle's worldline increases as the particle propagates from 
 emission 
to detection. But even $\kappa$-Minkowski provides no support for
the other key aspect of the intuition about (and the phenomenology on) spacetime
quantization provided by the heuristic arguments, which concerns an irreducible 
Planck-scale contribution to energy uncertainties.

In closing, it is perhaps useful to summarize what we feel are the key findings reported
 in this Letter.
We here performed the
 first ever constructive/deductive (no heuristics)
 derivation of the properties of fuzzy worldlines in a class of quantum spacetimes.
We established firmly that the phenomenological parametrization based 
exclusively on
the parameters $\alpha$ and $\beta$ (in the notation here adopted in the opening remarks),
which had been suggested by several heuristic arguments,
is at least not sufficiently general.
Specifically we established that this phenomenological parametrization is not applicable to
the two most studied noncommutative spacetimes,
the Moyal type and the $\kappa$-Minkowski type.
In the $\kappa$-Minkowski case we did find that, as argued by the heuristic arguments,
the fuzziness of a particle's worldline increases as the particle propagates from 
 emission 
to detection: combining (\ref{mainresult}) and (\ref{mainresultE})
one gets for $\kappa$-Minkowski
\begin{equation}
\delta {\cal A}_{[\ell]} \simeq \ell \, D \, \delta E
\label{mainresultF}
\end{equation}
where we only included the contribution growing with the propagation distance 
 and we used $D=a^0$ since the translation parameter $a^0$ connecting the observer at emission
 and the observer at detection is just the propagation distance
(any difference from this would only manifest itself at subleading orders in $\ell$).
The propagation-distance-dependent amplification of Planck-scale effects shown
in (\ref{mainresultE}) can provide a natural target for quantum-gravity phenomenology,
and now these plans can be pursued at a level that goes beyond heuristics.
But the phenomenology will need to adapt to the fact that within
our $\kappa$-Minkowski 
picture the $\delta E$ in (\ref{mainresultF}) receives no irreducible
Planck-scale contributions. 

The absence of irreducible
Planck-scale contributions to $\delta E$ also characterizes our Moyal-case results
(and in the Moyal case even the propagation-distance amplification of type (\ref{mainresultF}) 
is absent).
This is rather significant since the presence of irreducible Planck-scale contributions
to $\delta E$, of the type here noted in Eq.~(\ref{fuzzyMOMENTUM}),
was a key aspect of the conceptualization of spacetime fuzziness provided
by previous heuristic arguments, and a key ingredient of the associated
phenomenology developed in 
Refs.~\cite{ngPRL,steingbring,shviliBLURRING,tamburini,lieuHillman,ragazzoni}
and references therein.
The ``experimental bounds on spacetime fuzziness" derived in
Refs.~\cite{ngPRL,steingbring,shviliBLURRING,tamburini,lieuHillman,ragazzoni}
and references therein
are therefore evidently inapplicable to Moyal and $\kappa$-Minkowski noncommutativity.
And it would be reductive to view our study as a counter-example to a general feature:
this is the first time that the expectations of the relevant heuristic arguments
have been tested in actual formalizations of the notion of a quantum spacetime, 
and they were found to fail. There may well
be some quantum spacetimes where the relevant heuristic predictions do happen to apply, but there is
at present (with two actual spacetime pictures analyzed, and the heuristic predictions found to be
there inapplicable)
no reason to assume that those predictions will be generic.
It is still plausible that those bounds do apply for example
to the picture of quantum spacetime
emerging from Loop-Quantum-Gravity research~\cite{rovelliLivingReview}, 
and we feel that establishing rigorously
such a link (or the lack thereof) should be one of the next 
 main targets for this research programme.

\bigskip
\bigskip
\bigskip
$~$\\
{\it  This work was supported in part by a grant from the John 
Templeton Foundation. GR also acknowledges 
support by the Polish National Science Centre under the
agreement DEC-2011/02/A/ST2/00294.}

\end{document}